\documentclass[traditabstract,letter]{aa}

\usepackage{graphicx}
\usepackage{txfonts}
\usepackage{float}

\usepackage{natbib}
\bibpunct{(}{)}{;}{a}{}{,}

\usepackage{hyperref}
\hypersetup{colorlinks=true,
            citecolor=blue,
            linkcolor=blue}

\usepackage{color}

\newcommand{\orcid}[1]{\protect\href{https://orcid.org/#1}{\protect\includegraphics[width=8pt]{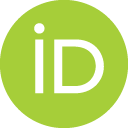}}}
 
\def \vv#1{\vec{#1}}
\def \be {\begin{equation}}
\def \ee {\end{equation}}

\def \m{(m_0+m_1)}
\def \C{C}
\def \G{G}
\def \w{\Omega}

\def \vr{\vv{r}}
\def \vw{\vv{\w}}
\def \vii{\vv{i}_0}
\def \vj{\vv{j}_0}
\def \vk{\vv{k}}
\def \vs{\vv{s}}
\def \vL{\vv{L}}
\def \vS{\vv{S}}

\def\H{\overline H}
\def\K{J}

\def\cte{\mathrm{const}}
\def\vp{\vv{p}}
\def\vK{\vv{\K}}

\def\ui{u}
\def\vi{v}
\def\xi{x_i}
\def\x{x_1}
\def\y{x_0}
\def\z{z}
\def\psh{\phi}
\def\dphi{\Delta \psh}
\def\zx{X(\x,\ui)}
\def\zxs{X^2(\x, \ui)}
\def\uic{\ui_c}
\def\deltai{\varepsilon}

\def\figpath{Figures/}
\def\bibpath{}
\def \llabel#1{\label{#1}}

\begin{document}

\title{Spin--spin coupling in stellar binaries}

\author{Alexandre C. M. Correia\inst{1,2}
\orcid{0000-0002-8946-8579}}

 \authorrunning{A. C. M. Correia}

 
\institute{
CFisUC, Departamento de F\'isica, Universidade de Coimbra, 3004-516 Coimbra, Portugal
\and 
LTE, Observatoire de Paris, Universit\'e PSL, Sorbonne Universit\'e, CNRS, 75014 Paris, France  
}

\date{Received 31 March 2026 / accepted 26 May 2026}

\abstract{Spin--orbit misalignment is increasingly observed in binary stars, but its origin remains uncertain. We study the secular spin dynamics of stellar binaries and derive the associated Cassini states. We isolate a spin--spin resonance that arises when the two stellar spin-precession frequencies become commensurate. This resonance can excite the obliquity of the secondary to high values even when the primary is only weakly tilted. Magnetic braking favors the formation of high-obliquity states, whereas tidal dissipation may suppress them. Spin--spin coupling may thus help explain the observed diversity of obliquities in young stellar binaries.}


\maketitle
\nolinenumbers


\section{Introduction}

Measurements of stellar obliquities in binaries have shown that spin--orbit misalignment is a real and possibly common feature of binary-star dynamics. 
The archetypal system DI~Herculis exhibits large stellar obliquities \citep{Albrecht_etal_2009}, while CV~Velorum provides another well-studied case with a significantly misaligned stellar spin axis \citep{Albrecht_etal_2014}.
Gaia-based population studies indicate that nonzero obliquity is not restricted to a few exceptional systems, and that a significant fraction of early-type binaries are misaligned \citep{Marcussen_etal_2024,Smith_etal_2024}.
This picture is further strengthened by obliquity measurements in short-period low-mass eclipsing binaries \citep{Spejcher_etal_2026}.
These results motivate renewed interest in dynamical pathways capable of creating, maintaining, or modifying stellar obliquities.

Most theoretical work considers primordial misalignment \citep[e.g.,][]{Lai_2014}, tertiary-driven evolution \citep[e.g.,][]{Anderson_etal_2017}, spin--orbit resonances involving circumbinary disks \citep[e.g.,][]{Anderson_Lai_2021}, or Cassini-state dynamics driven by nodal orbital precession \citep[e.g.,][]{Felce_Fuller_2023}.
However, these works typically treat the two stellar spins as effectively uncoupled to first order, so that one or both stars may evolve into spin--orbit equilibria independently. 
By contrast, the possibility of a spin--spin resonance, which arises when the two stellar spin-precession frequencies become commensurate \citep[e.g.,][]{Correia_etal_2016}, 
has received no dedicated attention. 

This question is especially relevant because the spin evolution of binary stars is shaped by both magnetic braking \citep[e.g.,][]{Skumanich_1972} and tidal torques \citep[e.g.,][]{Hut_1981}.
Their combined action has long been recognized as a key ingredient in binary evolution \citep[e.g.,][]{Hurley_etal_2002, Repetto_etal_2014} and has been revisited in both massive and low-mass binaries \citep{Song_etal_2018, Fleming_etal_2019}. 
Since the stellar spin-precession frequencies depend directly on the stellar rotation rates, their secular evolution can sweep binaries through the spin--spin commensurability. 
In this Letter, we isolate this resonance, characterize its dynamics, and assess its contribution to the observed diversity of spin--orbit angles in binary stars.

\section{Spin dynamics}

We consider a binary system composed of a primary and a secondary star, with masses $m_0$ and $m_1$, respectively, separated by the relative position vector $\vr$.
Each star of mass $m_i$ is modeled as an oblate ellipsoid with mean radius $R_i$, rotation period $P_i$, and angular velocity $\vw_i = \w_i \, \vs_i $, where $\w_i = 2 \pi / P_i$ and $\vs_i$ is the axis of maximum inertia (gyroscopic approximation) with moment of inertia $\C_i = \zeta_i m_i R_i^2$, where $\zeta_i$ is an internal structure constant.
The rotational angular momentum of each star is then
\be
\vS_i = S_i \, \vs_i = \C_i \, \w_i \, \vs_i \ , \llabel{141216a}
\ee 
and the orbital angular momentum is
\be
\vL 
= L \, \vk = \beta n a^2 \sqrt{1-e^2} \, \vk 
\ , \llabel{141216b}
\ee
where $\vk$ is the unit vector normal to the orbital plane, $a$ is the semi-major axis, $e$ is the eccentricity, $n = \sqrt{\mu / a^3}$ is the mean motion, and $ \beta = m_0 m_1 / \m $ is the reduced mass.

The Hamiltonian of the system is \citep[e.g.,][]{Smart_1953}: 
\be
H = \frac{\vp^2}{2 \beta}  
 -  \frac{\beta \mu}{r} \left[ 1 - \sum_{i=0,1} J_{2i} \left(\frac{R_i}{r}\right)^2 \!\! P_2 (\hat \vr \cdot \vs_i)
\right] 
+ \sum_{i=0,1} \frac{\vS_i^2}{2 \C_i} 
\ , \llabel{090514a}
\ee
where $\vp = \beta \dot \vr$, $\mu = \G \m $, $\G$ is the gravitational constant, $P_2(x) = (3x^2-1)/2$ is the Legendre polynomial of degree two, and $\hat \vr = \vr / r$ is the position unit vector; 
terms in $(R_i/r)^3$ have been neglected (quadrupolar approximation). 
For the gravity-field coefficients, we adopt \citep[e.g.,][]{Correia_Rodriguez_2013}
\be
J_{2i} = k_{2i} \frac{\w_i^2 R_i^3}{3 G m_i} \ ,
\ee
where $k_{2i}$ is the second Love number for potential. 

In general, the precession of the spin axes is much slower than the orbital motion.
Since we are only concerned with spin--spin interactions, we average the Hamiltonian (Eq.\,(\ref{090514a})) over one orbital period.
For the nonconstant terms, we obtain \citep[e.g.,][]{Goldreich_1966a, Boue_Laskar_2009} 
\be
\H = - \sum_{i=0,1} \frac{\alpha_i}{2} (\vs_i \cdot \vk)^2 \ , 
\quad \mathrm{with} \quad
\alpha_i =  \frac{3 \G m_0 m_1 J_{2i} R_i^2}{2 a^3 (1-e^2)^{3/2}}
\ . \llabel{141216c}
\ee

In the averaged problem, all quantities in the Hamiltonian (Eq.\,(\ref{141216c})) are constant, except the unit vectors $\vs_i$ and $\vk$, which can be related to the angular momentum components (Eqs.\,(\ref{141216a}), (\ref{141216b})).
The system’s secular evolution can therefore be described by the evolution of these components, derived from the Hamiltonian through Poisson brackets \citep[e.g.,][]{Dullin_2004},  
\be
\dot \vS_i = \{ \vS_i, \H \} = \frac{\partial \H}{\partial \vS_i} \times \vS_i = - \alpha_i \left( \vs_i \cdot \vk \right) \, \vk \times \vs_i
\ , \llabel{150821i}
\ee
\be
\dot \vL = \{ \vL, \H \} = \frac{\partial \H}{\partial \vL} \times \vL 
= -  \sum_{i=0,1} \alpha_i \left( \vs_i \cdot \vk \right) \, \vs_i \times \vk
\ . \llabel{150821h}
\ee
We verify that the norms of the individual angular momentum vectors are conserved, as is the total angular momentum,
\be
\vK = \vL + \vS_0 + \vS_1 = \cte \ .
\llabel{141216i}
\ee

Following \citet{Boue_Laskar_2009}, the equations of motion reduce to an integrable system.
They can be simplified if we consider only the relative position in space of the unit vectors $\vs_i$ and $\vk$, given by the direction cosines (Fig.\,\ref{figS})
\be
x_i = \cos \theta_i = \vs_i \cdot \vk \ , \quad \mathrm{and} \quad
z = \cos \deltai = \vs_0 \cdot \vs_1 \ .
\llabel{141216j}
\ee
With these notations, we can rewrite the Hamiltonian (Eq.\,(\ref{141216c})),
\be
H_0 = \alpha_0 \y^2 + \alpha_1 \x^2 = - 2 \H = \cte
\ , \llabel{141216n}
\ee
and the total angular momentum (Eq.\,(\ref{141216i})),
\be
x_\star = \y + q \, \x + \delta_1 \, z 
= \frac{\vK^2 - L^2 - S_0^2 - S_1^2}{2 S_0 L} = \cte  
\ , \llabel{141216o}
\ee
with $ q = S_1 / S_0 $ and $ \delta_i = S_i / L $.

\section{Solutions in the precession frame}
\llabel{newvars}

We now consider a frame ($\vii, \vj, \vk$) that follows the precession of $\vs_0$ about $\vk$, that is,
\be
\vii =  \frac{\vs_0 - \y \vk}{(1-\y^2)^{1/2}} \ , \quad \vj = \frac{\vk \times \vs_0}{(1-\y^2)^{1/2}} \ ,
\ee
where $\vj$ is along the line of the nodes between the orbital plane and the equatorial plane of the star with mass $m_0$.
In this frame, 
we can express $\vs_1 = (u,v,\x)$,
for which
\be
\ui = \vs_1 \cdot \vii = \frac{\z - \y \x}{(1-\y^2)^{1/2}} = \sin \theta_1 \cos \dphi \ , 
\llabel{141217a}
\ee
and
\be
\vi = \vs_1 \cdot \vj = \frac{\vk \cdot (\vs_0 \times \vs_1)}{(1-\y^2)^{1/2}} = \sin \theta_1 \sin \dphi \ ,
\llabel{141217b}
\ee
with $\dphi = \psh_1 - \psh_0$, where $\psh_i$ denotes the precession angle of $\vs_i$ measured along the orbital plane (Fig.\,\ref{figS}).
Because $\vs_1$ is a unit vector, we can also express
\be
\x = \vs_1 \cdot \vk = \left(1 - u^2 - v^2\right)^{1/2}  \ . 
\llabel{141217c}
\ee
Following \citet{Correia_2016}, we can get $\z$ from expression (\ref{141217a}),
\be
\z = \y \x + \ui \, (1-\y^2)^{1/2} \ , \llabel{141217d}
\ee
while $\y$ can be obtained by eliminating $\z$ in expression (\ref{141216o}),
\be
(1 + \delta_1 \x) \, \y + \delta_1 \ui \, (1-\y^2)^{1/2} = x_\star - q \, \x
\ , \llabel{141217e}
\ee
which can be explicitly solved for $\y$ as
\be
\y = \frac{(1 + \delta_1 \x) \zx - \delta_1 \ui \sqrt{1-\zxs}}{\sqrt {(1 + \delta_1 \x)^2+(\delta_1 \ui)^2 }}
\ , \llabel{141217f}
\ee
with
\be
\zx = \frac{x_\star - q \, \x}{\sqrt {(1 + \delta_1 \x)^2+(\delta_1 \ui)^2 }}
\ .
\llabel{141224a0}
\ee
Thus, $\y$ depends only on ($\x, \ui$), and hence on ($\ui,\vi$) (Eq.\,(\ref{141217c})); so does the Hamiltonian (Eq.\,(\ref{141216n})),
\be
H_0 = H_0 (\x, \ui,x_\star) = H_0 (\ui,\vi,x_\star) \ . \llabel{141217g}
\ee

In Fig.~\ref{TOI2119PS}, we show the secular trajectories for the spin of the secondary projected on the orbital plane in the TOI-2119 binary system (Table~\ref{Tab1}) when $P_0 \approx P_1=2$~day. 
These trajectories were obtained by plotting the level curves $H_0 (\ui,\vi,x_\star) =\cte$ (Eq.\,(\ref{141217g})), but they correspond to the integration of the secular equations (\ref{150821i}) and (\ref{150821h}) for different initial values of $\theta_1$.

\begin{figure}
\begin{center}
\includegraphics[width=0.91\columnwidth]{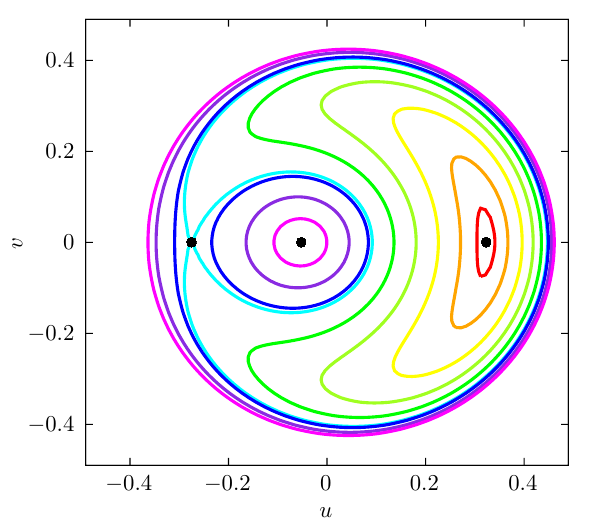} 
 \caption{Secular trajectories in the TOI-2119 system (Table~\ref{Tab1}) adopting $P_0=2.1$~day and $P_1=2.0$~day. We show the spin axis of the secondary, $\vs_1$, projected on the orbital plane. These trajectories are obtained by plotting the level curves $H_0 (\ui,\vi,x_\star) =\cte$ (Eq.\,(\ref{141217g})). The stationary solutions (Cassini states) are marked with a dot.  \llabel{TOI2119PS}  }
\end{center}
\end{figure}

The dynamics is akin to that of a second fundamental model for resonance \citep{Henrard_Lemaitre_1983}.
In astronomy, such behavior occurs for spin--orbit \citep[e.g.,][]{Ward_Hamilton_2004, Correia_2015} or mean--motion resonances \citep[e.g.,][]{Delisle_etal_2012, Petit_2021}.
Here, the resonant motion corresponds to a commensurability between the precession frequencies of the two stellar spins, $\dot \psh_0 \approx \dot \psh_1$, with (Eq.\,(\ref{150821i}))
\be
\dot \psh_i = - \alpha_i x_i / S_i 
\llabel{260326a} \ ; 
\ee
that is, it corresponds to a spin--spin resonance.

\section{Cassini states}

Cassini states are stationary equilibria of the spin axis \citep{Correia_2015} given by the extrema of the Hamiltonian (Eq.\,(\ref{141217g})),
\be
\frac{\partial H_0}{\partial \ui} = 0 \quad \mathrm{and} \quad \frac{\partial H_0}{\partial \vi} = 0 \ .
\llabel{150513a}
\ee 
Since $H_0 = H_0 (\x, \ui)$, we have
\be
\frac{\partial  H_0}{\partial \vi} = \frac{\partial  H_0}{\partial \x}\frac{\partial \x}{\partial \vi} = - \frac{\partial  H_0}{\partial \x}\frac{\vi}{\x} = 0 \ .
\llabel{150514a}
\ee 
We therefore conclude that $\vi = 0$ is always a possible equilibrium solution (equivalent to $\dphi = 0$ or $\pi$), where the unit vectors $\vs_0$, $\vs_1$, and $\vk$ remain coplanar.
We denote these coplanar states by $\uic = \pm \sin \theta_c $ (Eq.\,(\ref{141217a})),
which are given by (Eq.\,(\ref{141216n}))
\be
\left. \frac{\partial H_0}{\partial \ui} \right|_{\vi=0} = - 2 \alpha_1 \ui + 2 \alpha_0 \y \left. \frac{\partial \y}{\partial \ui} \right|_{\vi=0} = 0  \ .
\llabel{150513a}
\ee
Setting $\vi=0$ in $\x$ (Eq.\,(\ref{141217c})), we get $\x^c = \sqrt{1-\uic^2}$, and for $\y$ (Eq.\,(\ref{141217f})), we have $\y^c = \y(\x^c,\uic)$.
Equation~(\ref{150513a}) therefore provides an implicit condition for the coplanar states,
\be
\alpha_1 \uic = \alpha_0 \y^c \frac{\partial \y^c}{\partial \uic} 
\ , \llabel{150511a}
\ee
whose roots can be found in the interval $ \uic \in [-1,1]$ using numerical methods.
In Appendix~\ref{classicstates}, we provide approximate analytic expressions for these states when $S_1 \ll L$.

In the example shown in Fig.~\ref{TOI2119PS}, the stationary solutions correspond to three Cassini states, $\uic = -0.275$, $\uic = -0.0522$, and $\uic = 0.323$, equivalent to $\theta_1 \approx 16.0^\circ$, $\theta_1 \approx 3.0^\circ$, and $\theta_1 \approx 18.9^\circ$, respectively.
The smallest $\uic$ value corresponds to a hyperbolic unstable point, but the spin can be stabilized in the other two states.
The largest $\uic$ value lies within a libration region and thus corresponds to the resonant equilibrium.

In Fig.~\ref{TOI2119CS}, we plot the Cassini states as a function of the rotation period of the primary, $P_0$, for the TOI-2119 system (Table~\ref{Tab1}) with $P_1 = 2$~day.
For $P_0 \lesssim 2$~day, we observe that there is only one stable state at nearly zero obliquity.
For $P_0 \gtrsim 2$~day, the obliquity of the previous state increases to high values, while two additional states appear.
This is a consequence of the emergence of a separatrix that introduces a libration region around the high-obliquity state (Fig.~\ref{TOI2119PS}).
We then conclude that when the rotation period of the primary increases (e.g., through magnetic braking), the spins can be captured into resonance, leading to significant changes in the obliquity.

\begin{figure}
\begin{center}
\includegraphics[width=\columnwidth]{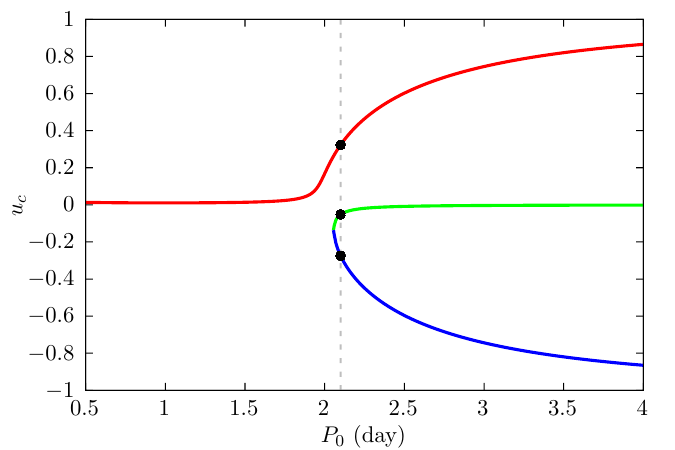} 
 \caption{Cassini states as a function of the rotation period of the primary, $P_0$, for the TOI-2119 system (Table~\ref{Tab1}) with $P_1 = 2$~day.
These equilibria are obtained by solving Eq.~(\ref{150511a}). 
The vertical dashed line corresponds to the configuration shown in Fig.~\ref{TOI2119PS}. \llabel{TOI2119CS}  }
\end{center}
\end{figure}

\section{Obliquity excitation}

So far, we have neglected magnetic braking and tidal torques, which modify the rotational angular momenta, $\vS_i$ (Eq.\,(\ref{141216a})).
These effects extract energy from the system, so the problem is no longer integrable and must be solved numerically.
In Appendices~\ref{MB:equations} and~\ref{tidal:equations}, we provide the secular equations for magnetic braking and tidal effects, respectively.

As an example, we consider the TOI-2119 binary system \citep{Doyle_etal_2025}, composed of an M-type main-sequence primary and a brown dwarf secondary (Table~\ref{Tab1}).
Observational studies of young brown dwarfs with masses in the range $0.02-0.08~m_\odot$ show rotation periods spanning $0.7-4.5$~day, with a median value of $1.9$~day, and a clear trend of longer periods at higher mass \citep{Scholz_etal_2018}.
Likewise, population studies of young low-mass stars in the range $0.15-1.5~m_\odot$ show a rotation period distribution of $0.2-30$~day, peaking near $2$~day, with M stars rotating faster on average than GK stars \citep{Rebull_etal_2018}. 
We therefore adopt an initial period of $P_0=1$~day for the primary and $P_1=2$~day for the secondary.
For the initial obliquities, we use the currently observed value $\theta_0 = 15.7^\circ$ for the primary and arbitrarily set $\theta_1 = 1^\circ$ for the secondary.

In Fig.~\ref{TOI2119SE} (left), we show the secular spin evolution of both stars under magnetic braking alone. 
The rotation period of the secondary is not expected to change significantly because it is a brown dwarf \citep[e.g.,][]{Zapatero-Osorio_etal_2006}. 
Thus, as the rotation period of the primary increases and reaches $P_0 \approx P_1 = 2$~day, the precession rates also become comparable, $\dot \psh_0 \approx \dot \psh_1$ (Eq.\,(\ref{260326a})), thereby modifying the phase space (Fig.~\ref{TOI2119CS}). 
The spin of the brown dwarf, initially circulating around the only existing Cassini state (red state in Fig.~\ref{TOI2119CS}), adiabatically follows this state, leading to a significant increase in its obliquity. 
In the absence of tides, this increase is permanent and can drive the obliquity to values close to $90^\circ$ for $P_0 \gg 2$~day.

\begin{figure*}
\begin{center}
\includegraphics[width=\columnwidth]{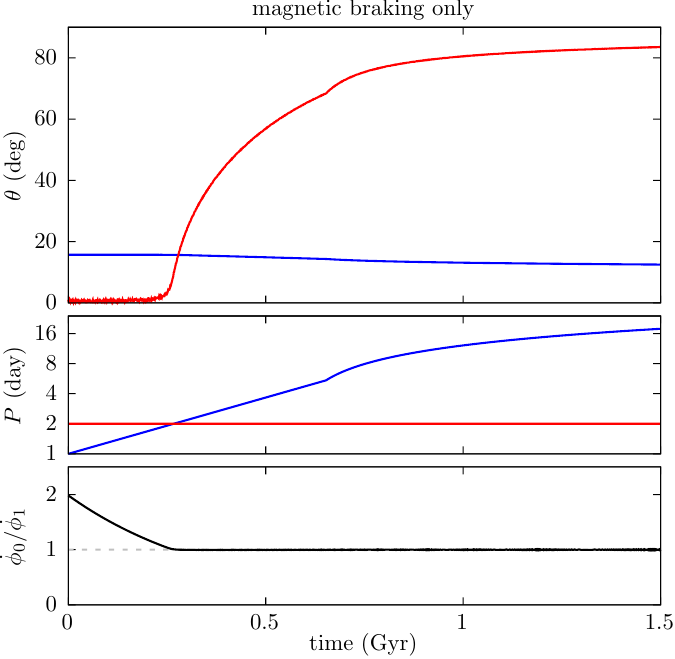} 
\includegraphics[width=\columnwidth]{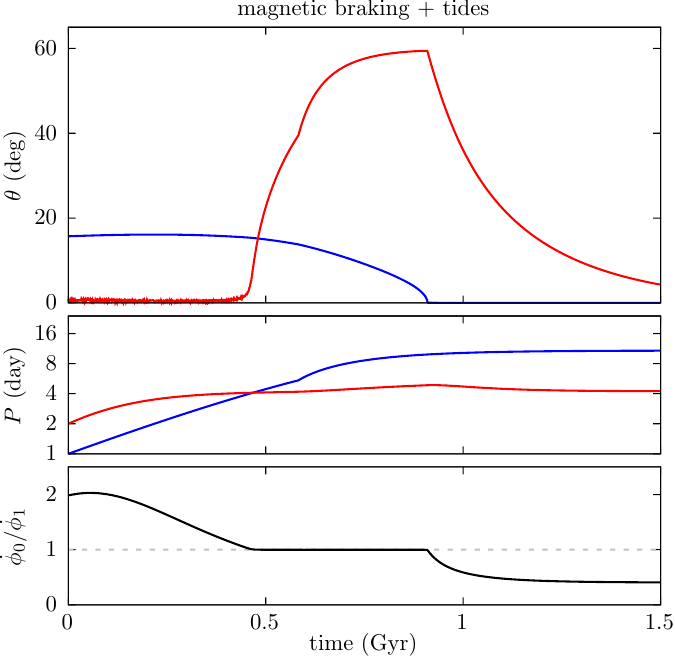} 
 \caption{Secular spin evolution of the TOI-2119 binary stars as a function of time. We show the obliquity (top), rotation period (middle), and precession frequency ratio (bottom). The primary and secondary spins are shown in blue and red, respectively. The left column shows the evolution under magnetic braking alone, while the right column also includes tidal effects for both stars.\llabel{TOI2119SE}  }
\end{center}
\end{figure*}

The obliquity of the primary, $\theta_0$, is also affected by the resonance crossing.
Because the total angular momentum is nearly conserved and $S_i \ll L$, we generally have (Eq.\,(\ref{141216o}))
\be
\cos \theta_0 + q \cos \theta_1 \approx \cte
\ . \llabel{260326b}
\ee
It follows that $\theta_0$ must decrease as $\theta_1$ increases. 
For the TOI-2119 system, the decrease in $\theta_0$ is small since $q \ll 1$ as well.

In Fig.~\ref{TOI2119SE} (right), we show the secular spin evolution including tides for both stars.
There are two main differences relative to the previous case.
First, tides also modify the secondary rotation period, driving it toward the pseudo-synchronous equilibrium at $P_1 = 4.2$~day (Eq.\,(\ref{090520a})) and thus delaying the resonance encounter.
Second, tides act directly on the obliquity, damping it to near zero \citep[e.g.,][]{Correia_etal_2016}.
As a result, although the secondary's obliquity, $\theta_1$, is still excited by the spin--spin resonance, the primary's obliquity, $\theta_0$, continues to decrease owing to tides.
Once $\theta_0$ is damped to nearly zero, the resonant equilibrium can no longer be maintained, and $\theta_1$ is also damped.

In Appendix~\ref{EBLM45}, we provide another example of a stellar binary with two main-sequence stars, EBLM~J2025-45.
In this case both stars undergo magnetic braking, which may give rise to complementary obliquity excitation behaviors in the absence of tides.
However, when tides are included, the system evolves in a way very similar to TOI-2119 (Figs.~\ref{EBLM45SE1} and~\ref{EBLM45SE2}).

\section{Discussion}

In this Letter, we studied the dynamics of spin--spin coupling in stellar binaries and its effect on their obliquities.
We showed that this mechanism can efficiently excite the obliquity of the secondary to high values even when the primary's obliquity is small.
However, the resonant equilibrium breaks down as the primary’s obliquity nears zero, owing to tidal dissipation.

Magnetic braking is the main driver of high-obliquity states, while tidal damping acts against it.
Although magnetic braking is well constrained by observational studies \citep[e.g.,][]{Gallet_Bouvier_2013}, tidal dissipation in stars relies mostly on theoretical models and can vary by orders of magnitude depending on stellar properties and age \citep[e.g.,][]{Ogilvie_2014, Mathis_2015}.
More efficient tidal damping may prevent high-obliquity states from developing, whereas weaker dissipation may allow such states to survive over the system’s lifetime.

The two examples shown, TOI-2119 and EBLM~J2025-45, are compact binaries ($a \lesssim 0.07$~au), where tides are more efficient. In wider binaries, magnetic braking is expected to dominate the evolution, allowing high-obliquity states to persist longer. However, the precession rates are then also slower (Eq.\,(\ref{260326a})), making the resonant excitation less efficient.

Spin--spin resonances are most likely to occur during the early stages of stellar evolution, when rotation is rapid and evolves quickly, but the resulting high-obliquity states may persist for Gyr.
In our examples, we adopted initial rotation periods near resonance, so that the resonance crossing occurs after a short time.
Larger initial period ratios may delay or prevent the onset of obliquity excitation.

Our results show that spin--spin coupling may play an important role in shaping the diversity of spin--orbit angles observed in young stellar binary systems.

\begin{acknowledgements}
We acknowledge support from FCT - Funda\c{c}\~ao para a Ci\^encia e a Tecnologia, I.P., Portugal, through the CFisUC project UID/04564/2025 (with DOI identifier 10.54499/UID/04564/2025).
\end{acknowledgements}

\bibliographystyle{aa}
\bibliography{\bibpath correia}


\begin{appendix}
\nolinenumbers

\section{Reference angles definition}

\begin{figure}[h!]
\begin{center}
\includegraphics[width=\columnwidth]{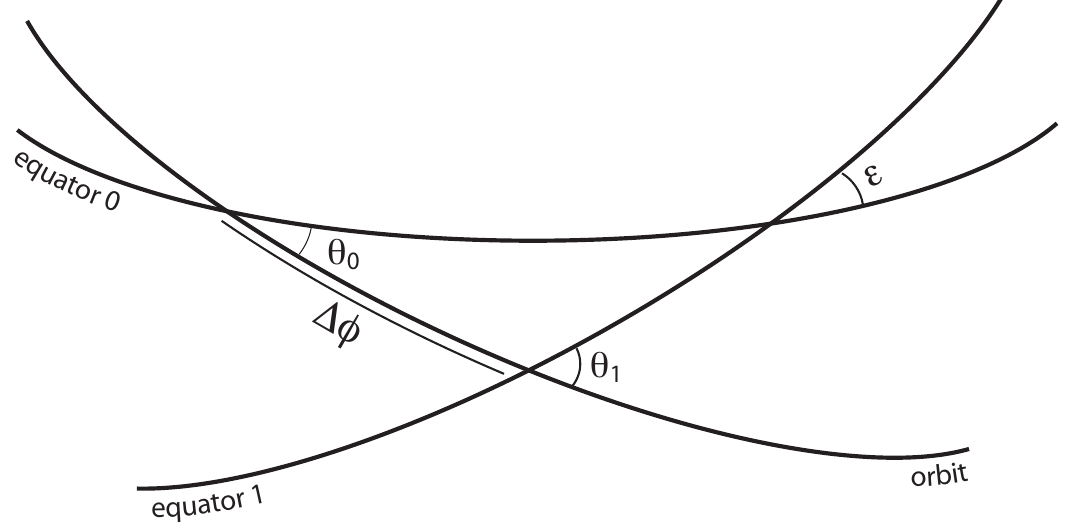} 
 \caption{Reference planes for the definition of the direction cosines.  \llabel{figS}  }
\end{center}
\end{figure}

\section{Parameters of the binary systems}

\begin{table}[h!]
\caption{Adopted parameters for the binary systems TOI-2119 \citep{Doyle_etal_2025} and EBLM~J2025-45 \citep{Spejcher_etal_2026}. The ages and the values of $\zeta_i$ and $k_{2i}$ are estimated from stellar evolution models \citep{Leconte_etal_2011a, Claret_2023, Sethi_etal_2026}.  \llabel{Tab1}} 
\centering
\begin{tabular}{|l|c|c|}
\hline
parameter & \phantom{45-}TOI-2119\phantom{-45} & EBLM~J2025-45 \\
\hline
$a$ [au] & $0.0611$ & $0.07061$ \\
$e$ & $0.3355$ & $0.12651$ \\
$m_0$ [$m_\odot$] & $0.53$ & $1.001$ \\
$m_1$ [$m_\odot$] & $0.06145$ & $0.224$ \\
$R_0$ [$R_\odot$] & $0.51$ & $1.053$ \\
$R_1$ [$R_\odot$] & $0.109$ & $0.249$ \\
$P_0$ [day] & $13.1$ & $6.43$ \\
$\theta_0$ [$^\circ$] & $15.7$ & $17.3$ \\
\hline
$\zeta_0$ & $0.178$ & $0.070$ \\
$\zeta_1$ & $0.210$ & $0.205$ \\
$k_{20}$ & $0.213$ & $0.028$ \\
$k_{21}$ & $0.340$ & $0.300$ \\
age [Gyr] & $0.7-5.1$ & $6.3-10$ \\
\hline
\end{tabular}
\end{table}

\section{Cassini states for $S_1 \ll L$}
\llabel{classicstates}

For binary systems, we usually have (Eqs.\,(\ref{141216a}),\,(\ref{141216b})),
\be
\delta_1 = \frac{S_1}{L} = \frac{\C_1 \w_1}{\beta n a^2 \sqrt{1-e^2}} 
\approx \zeta_1 \left(\frac{R_1}{a}\right)^2 \ll 1 
\ . \llabel{260328a}
\ee
Then, neglecting terms in $q \delta_1$ and $\delta_1^2$, we rewrite Eq.\,(\ref{141217f}) as
\be
\y \approx x_\star - q \, \x - \delta_1 \left( x_\star \x + \ui \, (1-x_\star^2)^{1/2} \right)
\ , \llabel{260330b}
\ee
and so
\be
\left. \frac{\partial \y}{\partial \ui} \right|_{\vi=0} \approx 
q \, \ui/\x + \delta_1 \left( x_\star \ui / \x - (1-x_\star^2)^{1/2} \right) 
\ . \llabel{260330c}
\ee
Replacing in Eq.\,(\ref{150511a}), we get for the Cassini states
\be
\alpha_1 \uic \x^c \approx \alpha_0 ( x_\star - q \, \x^c ) \left[ q \, \uic + \delta_1 \left( x_\star \uic  - \x^c (1-x_\star^2)^{1/2} \right) \right]
\ , \llabel{260330d}
\ee
where $\x^c = \sqrt{1-\uic^2}$.
This is a quartic equation in $\uic$, which may have up to four real roots.
To find them, we can use the trigonometric relations,
\be
\uic = \sin \theta_c \ , \quad \x^c = \cos \theta_c \ , \quad \mathrm{and} \quad x_\star = \cos \theta_\star \ ,
\ee
where $\theta_\star$ corresponds approximately to the initial $\theta_0$ (Eq.\,(\ref{141216o})).
Replacing in Eq.\,(\ref{260330d}), we get for the Cassini states, 
\be
\frac{\alpha_1}{\alpha_0} \sin \theta_c \cos \theta_c \approx ( \cos \theta_\star - q \cos \theta_c ) \Big[ q \sin \theta_c + \delta_1 \sin (\theta_c - \theta_\star) \Big]
\ , \llabel{260330e}
\ee
whose approximate roots are
\be
\theta_c \approx \arctan \left( \frac{\delta_1 \sin \theta_\star (q \pm \cos \theta_\star) }{\alpha_1/\alpha_0 + q^2 \pm q \cos \theta_\star} \right) 
\ , \llabel{260330f}
\ee
and
\be
\theta_c \approx \pm \arccos \left( \frac{q \cos \theta_\star}{\alpha_1/\alpha_0 + q^2} \right) 
\ . \llabel{260330g}
\ee
We also conclude that the high-obliquity states (Eq.\,(\ref{260330g})) are only possible when 
\be
 \alpha_1/\alpha_0 \ge q \left( \cos \theta_\star - q \right)
\ . \llabel{260331a}
\ee

\section{Magnetic braking}
\llabel{MB:equations}

A star's magnetic field couples to its surrounding ionized wind.
As the wind escapes, it extracts angular momentum from the star, causing its rotation to slow down over time.
This process, known as magnetic braking, is particularly important in low- and intermediate-mass stars, which possess substantial convective envelopes and are therefore able to sustain efficient magnetic dynamos.
The efficiency of magnetic braking is commonly parametrized in terms of the convective turnover timescale, $\tau_i$, which depends on stellar mass\footnote{The convective turnover timescale is the time required for a convective element to traverse the stellar convection zone.}.

For $0.08 < m_i / m_\odot < 1.36$, we adopt \citep{Wright_etal_2018}
\be
\log_{10} \tau_i = 2.33 - 1.50 \, \left(\frac{m_i}{m_\odot}\right) + 0.31 \, \left(\frac{m_i}{m_\odot}\right)^2 \ , \llabel{251021a}
\ee
where $\tau_i$ is expressed in days.
For $m_i > 1.36 \ m_\odot$, the convective turnover timescale decreases steeply.
Stellar evolution models indicate that it drops by approximately two orders of magnitude for $m_i \sim 2$~$ m_\odot $ \citep[e.g.,][]{Amard_etal_2019}.

The efficiency of magnetic braking also depends sensitively on the Rossby number,
${\cal R}_i = P_i / \tau_i $.
When the Rossby number falls below a critical threshold, ${\cal R}_S$, several indicators of magnetic activity appear to saturate, reaching an approximately constant maximum value that is largely independent of ${\cal R}_i$.
Following \citet{Matt_etal_2015}, we write ${\cal R}_S = {\cal R}_\odot / \chi$, where ${\cal R}_\odot$ is the solar Rossby number and $\chi \approx 10$.

The magnetic-braking torque can then be written, for rapidly rotating stars in the saturated regime (${\cal R}_i < {\cal R}_S$), as
\be
\dot \vS_i  = - \dot S_\odot \left( \frac{R_i}{R_\odot} \right)^{3.1} \left( \frac{m_i}{m_\odot} \right)^{0.5} \chi^2 \left( \frac{\w_i}{\w_\odot} \right) \, \vs_i \ ,
\llabel{251028s}
\ee
whereas for more slowly rotating stars in the unsaturated regime (${\cal R}_i > {\cal R}_S$), we use
\be
\dot \vS_i  = - \dot S_\odot  \left( \frac{R_i}{R_\odot} \right)^{3.1} \left( \frac{m_i}{m_\odot} \right)^{0.5} \left( \frac{\tau_i}{\tau_\odot} \right)^2 \left( \frac{\w_i}{\w_\odot} \right)^3 \vs_i \ ,
\llabel{251028u}
\ee
where $\tau_\odot \approx 12.9$~days is the solar convective turnover timescale, 
and $ \dot S_\odot  \approx 6.3 \times 10^{23}$~J $\approx 1.4 \times 10^{-14}$~$m_\odot$\,au$^2$\,yr$^{-2}$ is a constant calibrated against observations \citep{Matt_etal_2019}.

\section{Tidal evolution}

\llabel{tidal:equations}

Tidal effects arise from the differential and inelastic deformation of each star under the gravitational forcing of its companion. Because the stars are not perfectly rigid, this perturbation produces a distortion that gives rise to a tidal bulge.
The dissipation of mechanical energy inside the star introduces a time delay, $\Delta t$, between the initial perturbation and the resulting deformation.
As a consequence, the companion exerts a torque on the tidal bulge, which modifies the spin and the orbit.

The exact dependence of $\Delta t$ on the frequency of the tidal perturbation is unknown, because it depends on stellar properties and age \citep[e.g.,][]{Ogilvie_2014}.
For simplicity, we adopt here a model with constant $\Delta t$, which can be made linear \citep{Singer_1968, Hut_1981}.
The equations of motion are \citep[e.g.,][]{Correia_2009}
\be
\dot \vS_i = - n K_i \left[ f_1(e) \frac{\vs_i + \cos \theta_i \, \vk}{2} \frac{\w_i}{n} - f_2(e) \, \vk \right] \ , 
\llabel{090514k}
\ee
\be
\dot \vL = \sum_{i=0,1} n K_i  \left[ f_1(e) \frac{\vs_i + \cos \theta_i \, \vk}{2} \frac{\w_i}{n} - f_2(e) \, \vk \right] 
\ , \llabel{260328c}
\ee
\be
\dot a = \sum_{i=0,1} \frac{2 K_i}{\beta a} \left[ f_2(e) \cos \theta_i \frac{\w_i}{n} - f_3(e)
\right] 
\ , \llabel{260328b}
\ee
with 
\be
K_i =  \frac{3 G R_i^5 m_{(1-i)}^2 }{a^6} \, k_{2 i} \Delta t_i 
\ , \llabel{090514m}
\ee
and
$ e^2  = 1 - L^2/(\beta^2 \mu a)$,
$ f_1(e) = (1 + 3e^2 + 3e^4/8) / (1-e^2)^{9/2} $, 
$ f_2(e) = (1 + 15e^2/2 + 45e^4/8 + 5e^6/16) / (1-e^2)^{6} $,
$ f_3(e) = (1 + 31e^2/2 + 255e^4/8 + 185e^6/16 + 25e^8/64) / (1-e^2)^{15/2} $.

Tidal dissipation in stars can vary by several orders of magnitude \citep[e.g.,][]{Mathis_2015}.
Observational constraints on the mean tidal quality factor in eclipsing binaries suggest $10^5 \lesssim Q \lesssim 10^7$ \citep[e.g.,][]{Patel_etal_2023}, while the distribution of close-in exoplanets around Sun-mass stars gives $ Q \gtrsim 10^7$ \citep[e.g.,][]{Penev_etal_2012}.
We therefore assume here $\Delta t_i = 0.05$~sec for $m_i < 0.8$~$m_\odot$ and $\Delta t_i = 0.005$~sec for $m_i > 0.8$~$m_\odot$, which correspond to $Q_i \approx 10^6$ and $Q_i \approx 10^7$, respectively, using $Q_i^{-1} = n \Delta t_i $.

The equilibrium spin is reached when $\dot \vS_i = 0$ (Eq.\,(\ref{090514k})), that is, for \citep[e.g.,][]{Correia_2009}
\be
\frac{\w_i}{n} = \frac{f_2(e)}{f_1(e)} \, \frac{2 \cos \theta_i}{1 + \cos^2 \theta_i} \ , \llabel{090520a}
\ee
which is also known as pseudo-synchronous rotation.
The corresponding timescale is $\tau_{\rm spin} \sim C_i / K_i $, whereas for the circularization of the orbit we get $\tau_{\rm orb} \sim \beta a^2 / K_i $.
For both systems in Table~\ref{Tab1}, we obtain $\tau_{\rm spin} \sim 1$~Gyr for the primary, $\tau_{\rm spin} \sim 100$~Myr for the secondary, and $\tau_{\rm orb} \sim 100$~Gyr.

\section{Application to EBLM~J2025-45}

\llabel{EBLM45}

As a complementary example of obliquity excitation, we consider here the EBLM~J2025-45 binary system \citep{Spejcher_etal_2026}, which is composed of a Sun-like G-type primary and an M-type secondary main-sequence stars (Table~\ref{Tab1}).

For guidance, in Fig.~\ref{EBLM45CS}, we plot the Cassini states for the EBLM~J2025-45 system as a function of the rotation period ratio, $P_0 / P_1$.
This picture is similar to that of TOI-2119 (Fig.~\ref{TOI2119CS}).
For $P_0 / P_1 \lesssim 1$, there is only one stable state at nearly zero obliquity.
For $P_0 / P_1 \gtrsim 1$, the obliquity of the original state increases to high values, while two additional states appear.

\begin{figure}
\begin{center}
\includegraphics[width=\columnwidth]{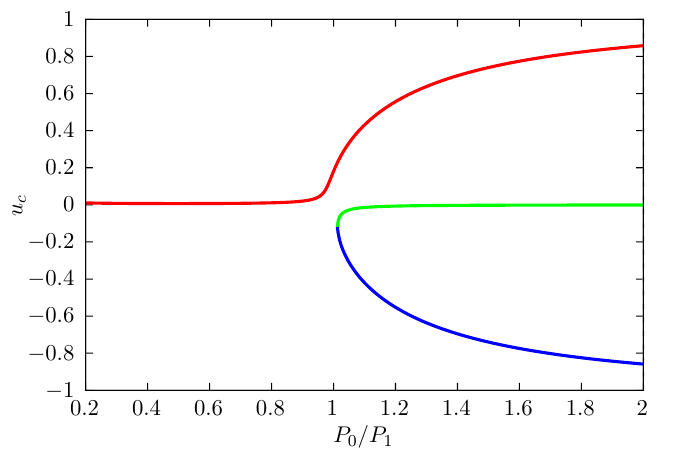} 
 \caption{Cassini states as a function of the rotation period ratio, $P_0/P_1$, for the EBLM~J2025-45 system (Table~\ref{Tab1}) with $P_1 = 2$~day.
These equilibria are obtained by solving Eq.\,(\ref{150511a}). \llabel{EBLM45CS}  }
\end{center}
\end{figure}

Observational studies of young low-mass stars in the range $0.15-1.5$~$m_\odot$ show rotation periods spanning $0.2-30$~day, peaking near $2$~day \citep{Rebull_etal_2018}.
In a first experiment, we thus adopt initial rotation periods of $P_0=1.5$~day for the primary and $P_1=2.0$~day for the secondary (Fig.~\ref{EBLM45SE1}). 
In a second experiment, we swap these values, adopting $P_0=2.0$~day for the primary and $P_1=1.5$~day for the secondary (Fig.~\ref{EBLM45SE2}).
For the initial obliquities, we use the currently observed value $\theta_0 = 17.3^\circ$ for the primary and arbitrarily set $\theta_1 = 5^\circ$ for the secondary.

In Fig.~\ref{EBLM45SE1} (left), we show the secular spin evolution of both stars under magnetic braking alone. 
We initially have $P_0 < P_1$ and the rotation period of the secondary, $P_1$, varies more slowly than that of the primary, $P_0$.
As a result, the rotation periods of both stars become equal after a short time, bringing the system into resonance. 
The obliquity of the secondary, $\theta_1$, then increases, following the high-obliquity Cassini state (red state in Fig.~\ref{EBLM45CS}), while the obliquity of the primary, $\theta_0$, decreases (Eq.\,(\ref{260326b})). 
The obliquity of the secondary peaks around $60^\circ$, corresponding to a maximum period ratio $P_0 / P_1$. As the G-type star spins down, it enters the unsaturated regime, where magnetic braking is less efficient (Eq.\,(\ref{251028u})). 
At that point, the ratio $P_0 / P_1$ begins to decrease, also leading to a decrease in $\theta_1$ (red state in Fig.~\ref{EBLM45CS}). 
When $P_0 / P_1 < 1$, the system moves out of resonance, since there is only one Cassini state left.

In Fig.~\ref{EBLM45SE1} (right), we show the secular  spin evolution including tides for both stars. Tides raised in the secondary are very efficient, so its rotation period increases much faster than in the absence of tides. 
Yet, after some time, the secondary's rotation period stabilizes at $P_1 = 6.2$~day, near the pseudo-synchronous equilibrium (Eq.\,(\ref{090520a})), allowing it to be caught up with the primary's rotation period. 
Then, the system enters in resonance, and $\theta_1$ is excited, while $\theta_0$ decreases owing to tides. Once $\theta_0$ is damped to nearly zero, the resonant equilibrium can no longer be maintained, and $\theta_1$ is also damped in a way very similar to that observed for the TOI-2119 system (Fig.~\ref{TOI2119SE}).

\begin{figure*}
\begin{center}
\includegraphics[width=0.989\columnwidth]{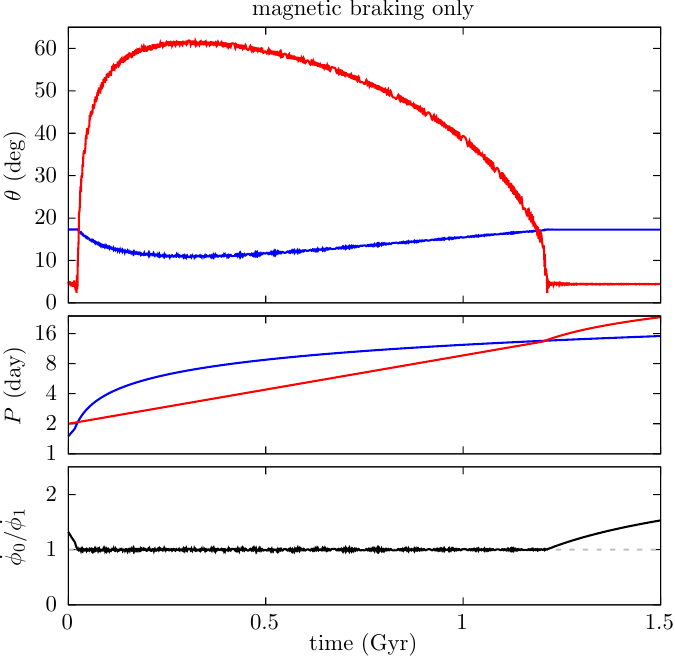} 
\includegraphics[width=0.989\columnwidth]{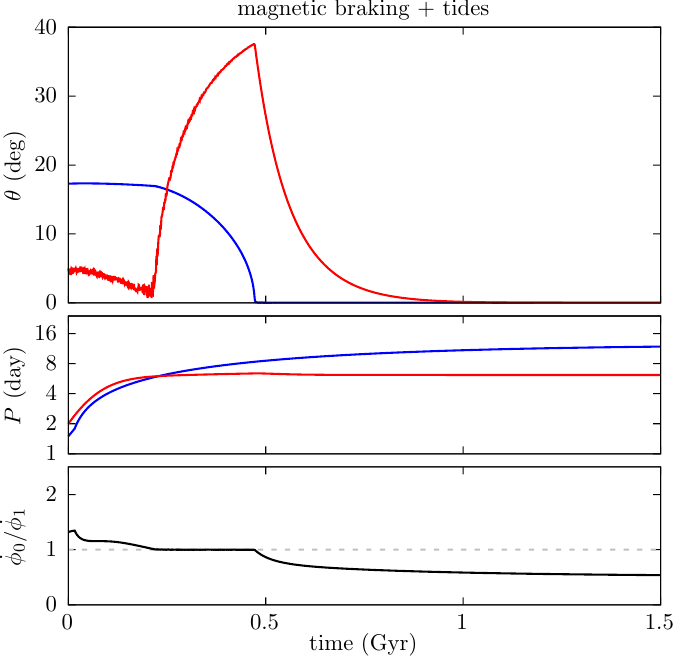} 
 \caption{Secular spin evolution of the EBLM~J2025-45 binary stars as a function of time, starting with $P_0=1.5$~day and $P_1=2.0$~day. We show the obliquity (top), rotation period (middle), and precession frequency ratio (bottom). The primary and secondary spins are shown in blue and red, respectively. The left column shows the evolution under magnetic braking alone, while the right column also includes tidal effects for both stars. \llabel{EBLM45SE1}  }
\end{center}
\end{figure*}

\begin{figure*}
\begin{center}
\includegraphics[width=0.989\columnwidth]{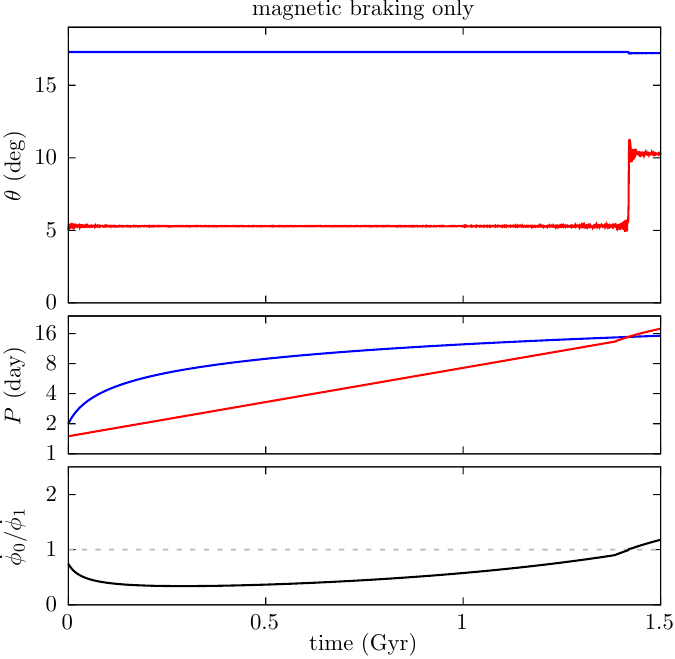} 
\includegraphics[width=0.989\columnwidth]{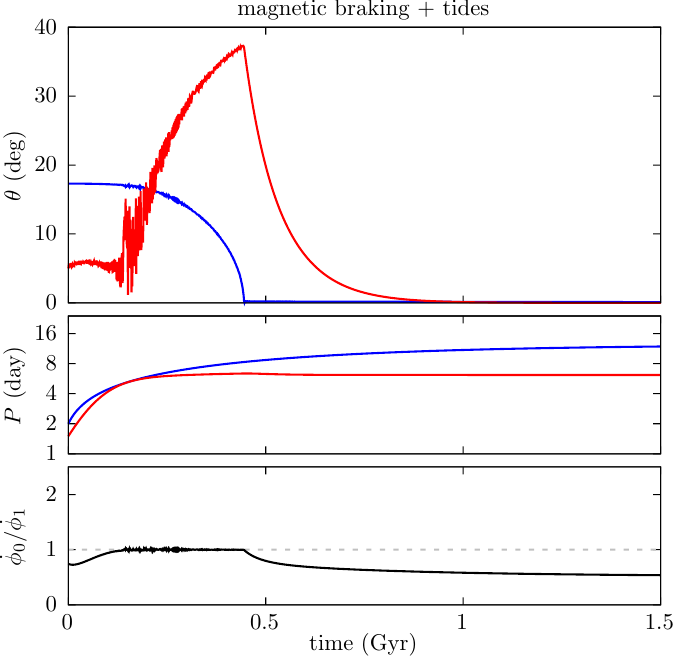} 
 \caption{Secular spin evolution of the EBLM~J2025-45 binary stars as a function of time, starting with $P_0=2.0$~day and $P_1=1.5$~day. We show the obliquity (top), rotation period (middle), and precession frequency ratio (bottom). The primary and secondary spins are shown in blue and red, respectively. The left column shows the evolution under magnetic braking alone, while the right column also includes tidal effects for both stars. \llabel{EBLM45SE2}  }
\end{center}
\end{figure*}

In Fig.~\ref{EBLM45SE2} (left), we show the secular spin evolution of both stars under magnetic braking alone.
In this case, we initially have $P_0 > P_1$, and so there is no resonance crossing at early times.
However, as the G-type star enters the unsaturated regime, the evolution of its rotation period slows down, allowing $P_1$ to catch up with $P_0$.
At that point, the system crosses the resonance, but with a decreasing ratio $P_0 / P_1$.
That is, before the resonant encounter, the system circulates around a low-obliquity Cassini state (green state in Fig.~\ref{EBLM45CS}). Then, this state disappears, and the system must circulate around the only remaining Cassini state (red state in Fig.~\ref{EBLM45CS}).
Nevertheless, although in this case resonant capture is not possible, the obliquity of the secondary is still slightly excited by about $5^\circ$.

In Fig.~\ref{EBLM45SE2} (right), we show the secular spin evolution including tides for both stars. Although we initially have $P_0 > P_1$, tides raised in the secondary allow $P_1$ to increase faster than $P_0$. The system then crosses the resonance with an increasing ratio $P_0/P_1$, allowing resonant capture and subsequent excitation of $\theta_1$. Once $\theta_0$ is damped to nearly zero owing to tides, the resonant equilibrium is broken, and $\theta_1$ is also damped in a way very similar to that observed for the TOI-2119 system (Fig.~\ref{TOI2119SE}).

\end{appendix}

\end{document}